\documentclass{MYws-ijmpa}

\begin{document}

\markboth{B. Loiseau, S. Wycech}
{Antiproton-proton resonant like channels
in $J/\psi\to\gamma p\bar p$ decays}

\catchline{}{}{}{}{}
%
%%%%%%%%%%%%%%%%%%%%%%%%%%%%%%%%%%%%%%%%%%%%%%%%%%%%%%%%%%%%%%%%%%%%

\title{ANTIPROTON-PROTON RESONANT LIKE CHANNELS\\
IN $J/\psi\to\gamma p\bar p$ DECAYS\footnote{Invited talk at the $10^{\mathrm{th}}$ International Symposium on Meson-Nucleon Physics and the Structure of the Nucleon (MENU 2004), Beijing, China, August 29 - September 4, 2004 - LPNHE 2004-12}
}

\author{\footnotesize B. LOISEAU}

\address{LPNHE\footnote{Unit\'e de Recherche des Universit\'es Paris 6 et Paris 7, associ\'e au CNRS}, Groupe Th\'eorie, Universit\'e P. \& M. Curie,\\
 4 Pl. Jussieu, F-75252 Paris, France
}

\author{S. WYCECH}

\address{Soltan Institute for Nuclear Studies,\\
PL - 00681 Warszawa, Poland
}

\maketitle

%\pub{Received (Day Month Year)}{Revised (Day Month Year)}

\begin{abstract}
The BES collaboration has recently observed a strong enhancement close to the proton-antiproton, $p\bar p$ threshold in the $J/\psi$ decays into $\gamma p\bar p$.
Such a structure  can be explained by a traditional nucleon-antinucleon, $N\bar N$, model. 
The near threshold $^{11}S_0$ bound state and/or the well-established $^{13}P_0$ resonant state found in this $N\bar N$ interaction can adequately describe the BES data.


\end{abstract}

\section{Introduction}  %) A SECTION HEADING

Existence of near threshold bound states or resonances in nucleon-antinucleon, $N\bar N$, interaction is a challenging matter\cite{klem02}.
Low-energy scattering could indicate the presence of such structures  by determining the scattering lengths for $^{2I+1,2S+1}L_J$ states.
Here $I$ denotes the isospin (0 or 1), $S$ the spin (0 or 1), $L$ the angular momentum and $J$ the total angular momentum.
An alternative is to use formation experiments. 
At the Beijing electron-positron collider, the BES collaboration has observed a resonant-like behavior in the $p\bar p$ invariant mass spectrum from radiative $J/\psi\to\gamma p\bar p$ decays\cite{bai03}.
The present work studies the physics of slow $p\bar p$ pairs produced in $J/\psi$ decays,
using $J^{PC}$ conservation, $P$ being the parity and $C$ the charge conjugation. 
Here we rely on the Paris $N\bar N$ potential model.

\section{Close to Threshold Proton-Antiproton Final State Model}
\subsection{The low-energy nucleon-antinucleon interaction}

The Paris $N\bar N$ interaction is built up from a state dependent optical potential.
The long range, $r>1$ fm, real part is obtained by G-parity transformation of the Paris $NN$ potential, the two-pion exchange of which is calculated via dispersion relations from pion-nucleon scattering data.
The short ranges, $r<1$ fm, real part and absorptive part, with a form suggested by calculation of $N\bar N$ annihilation into two mesons or resonances, are both determined through fit to the $\bar NN$ data.
In the different versions, the short range  parameters are readjusted by fitting to new data.
The Paris 82 potential\cite{cote82},  fitted to pre-LEAR (CERN) data, mainly elastic $\bar pp$ (isospin 1 + isospin 0) data , has a $\chi^2$/data of 2.8 for 915 data.
The Paris 94 potential\cite{pign94}  uses LEAR data, in particular $\bar pp\to\bar nn\ (I=1-I=0)$ data, and has a $\chi^2$/data of 2.46 for 3295 data.
In the Paris 99 version\cite{elbe99},  more recent LEAR data, in particular for $\bar pp\to\bar nn$, were used leading to a $\chi^2$/data of 2.95 for 3814 data.
The Paris 04 model\cite{lac04} is constrained by fitting to the 1999 data plus the scattering lengths extracted from antiprotonic hydrogen and deuterium data\cite{aug99} and to the total $\bar np$ cross-section\cite{iazz00}.
It has $\chi^2$/data = 3.19 for 3934 data.

\subsection{Allowed slow $p\bar p$ final states}

The $J^{PC}$ conservation $(J^{PC}=1^{--}\mbox{ for }J/\psi)$ limits the number of slow $p\bar p$ final states. These correspond to pairs of small $M_{p\bar p}-2m_p$ with $M_{p\bar p}$ being the invariant $p\bar p$ mass and $m_p$ the proton mass.
The allowed states are listed in Table 1.
Some  two-particle analogues\cite{pilk67} are listed in the second column.
The last column indicates the relative angular momentum between $\gamma$ or $\pi$ and the $p\bar p$ pair $h$.
The BES experiment\cite{bai03} angular distribution favors a pseudoscalar $^1S_0$ or a scalar $^3P_0$ $h$ final state.

\begin{table}[h]
\tbl{The slow $p\bar p$ pairs states permitted in the radiative $J/\psi\to\gamma p\bar p$.}
{\begin{tabular}{lccccc}\toprule
decay mode &   analogue                      &  $J^{PC}[\gamma \mbox{ or } \pi] $    &  $J^{PC}[p \bar{p} ]$   & $h(p\bar p)$ & relative $\ell$ \\
\colrule
  $\gamma  p \bar{p} (^1S_0)$ & $\gamma  \eta(1444)  $  & $ 1^{--}$    &  $0^{-+}$  &  pseudoscalar  &  1              \\
  $\gamma  p \bar{p} (^3P_0)$ & $\gamma  f_0(1710)   $  &  $ 1^{--} $  &  $ 0^{++} $  & scalar     &  0              \\
  $\gamma  p \bar{p} (^3P_1)$ & $\gamma  f_1(1285)   $  &  $ 1^{--} $  &  $ 1^{++} $  &  pseudovector    &  0              \\
\botrule
\end{tabular}}
\end{table}

\vspace*{-0.5cm}
\subsection{Specific final-state interaction model}

The transition amplitude from a channel $i$ to a channel $f$, in a multichannel system at low energy described by a S-wave K matrix, can be written as
$T_{if}=A_{if}(1+iq_fA_{ff})^{-1}$.
Here $A_{if}$ is a transition length, $A_{ff}$ the scattering length in the channel $f$ and $q_f$ the momentum in this channel\cite{pilk67}.
The  $f$ channel scattering amplitude can also be expressed as 
$T_{ff}=A_{ff}(1+iq_fA_{ff})^{-1}$.
For a P wave close to threshold,
$A_{ff}=A_{ff}^Pq_f^2$ and $A_{if}=A_{if}^Pq_f$
where $A_{ff}^P$ is the scattering volume. Up to terms in $q_f^2$ one has
$T_{if}=(A_{if}/A_{ff})T_{ff}=CT_{ff}/q_f^L=Ct_L$.
The quantity 
$C=A_{if}q_f^L/A_{ff}$
represents the unknown formation amplitude and
$\vert t_L\vert^2=\vert T_{ff}/q_f^L\vert^2$
is the final state interaction factor in a given $p\bar p$ partial wave.
In terms of the phase shifts $\delta_L$ and inelasticities $\eta_L$ of a given $N\bar N$ interaction one has
$t_L=\left(\eta_Le^{2i\delta_L}-1\right)/(2iq_f^{2L+1}).$
The function $C$ is parametrized by
$\vert C(x)\vert^2=q_f(c_0+c_1x)$
where 
$x=M_{p\bar p}-2m_p$ 
and
$q_f=[x(m_p+x/4)]^{1/2}$.

\section{Results and Conclusions}

The final state interaction factors 
$\vert t_L\vert^2$ for the $^1S_0$ and $^3P_0$ states and for the different versions of the Paris $N\bar N$ are compared to the BES data \cite{bai03} in Figs. 1 and 2.
The $c_0$ and $c_1$ parameters are determined by requiring
$\vert T_{if}\vert^2$
of Paris 04 to be close to the events distribution as given in Fig. 3 of Ref. 2 at $x=7$ MeV and 
$x=66.2$~MeV.
For $^1S_0$,  $c_0=1.18599$, $c_1=0.00299$ and 
for $^3P_0$,  $c_0=2.5206$, $c_1=0.0269$.
As seen in Fig. 1, the data is well reproduced by the Paris 04 $N\bar N$ interaction.
This interaction has a $^{11}S_0$ bound state  located at $x=-4.8$ MeV and with a width $\Gamma$ of 52.5 MeV.
Paris 99 has also a bound state at $x=-69$ MeV with $\Gamma=46$ MeV.
There are no bound states for Paris 94 or Paris 82.
All Paris models have a $^{13}P_0$ resonance of mass $\sim$ 1876 MeV and $\Gamma\sim 10$ MeV. They all reproduce the near threshold BES enhancement as seen in Fig. 2.

In conclusion, the near  threshold $p\bar p$ enhancement seen in BES collaboration\cite{bai03} can find a natural explanation from a traditional model of $\bar pp$ interaction.
The  $^{11}S_0$ bound state\cite{lac04}  needs confirmation.
The well established $^{13}P_0$ resonance originates from the strong attraction of the one-pion exchange\cite{zou04}.
Each of these states gives a reasonable representation of the BES radiative 
$J/\psi\to\gamma p\bar p$ decay data.
They correspond to the S Ñ or P Ñ wave Breit Wigner resonance functions considered by BES collaboration in their fit to the data\cite{bai03}.

\begin{figure}[h]
\begin{center}
\includegraphics*[scale=0.5,angle=90]{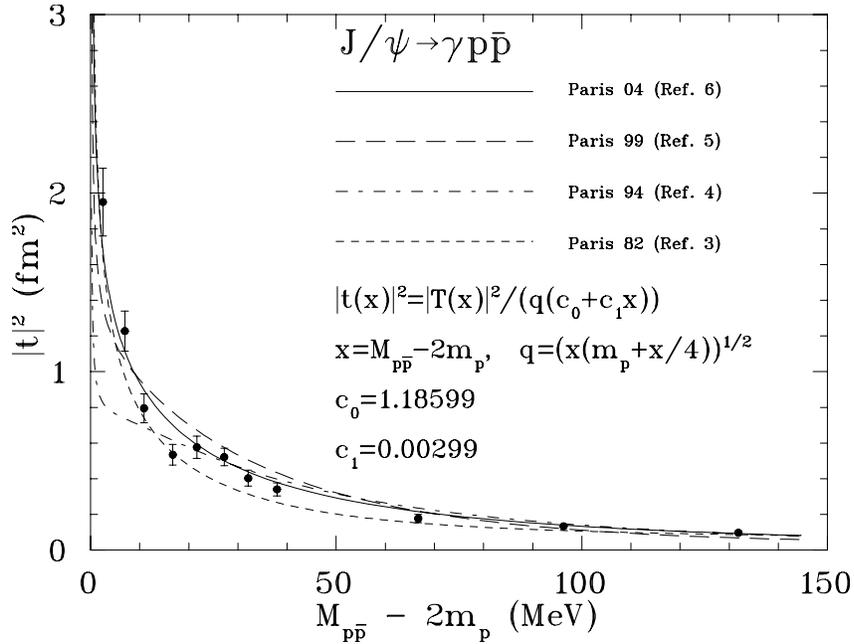}
\caption{The $^1S_0$ final state factor compared to BES data\protect \cite{bai03}}
\end{center}
\end{figure}

\begin{figure}[t]
\begin{center}
\includegraphics*[scale=0.45,angle=90]{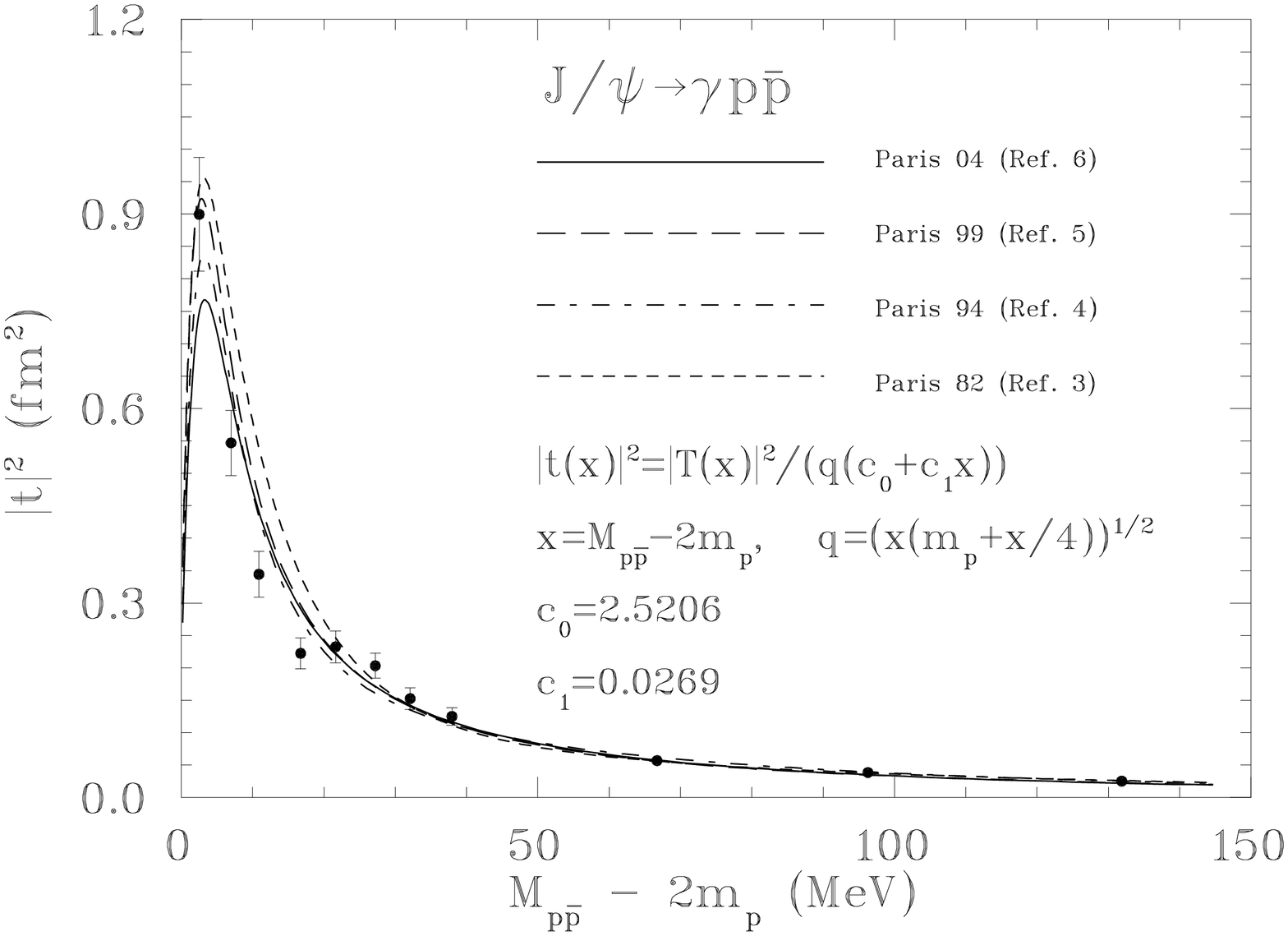}
\caption{The $^3P_0$ final state factor compared to BES data\protect \cite{bai03}}
\end{center}
\end{figure}

\end{document}